\documentclass[3p,times]{elsarticle}

\usepackage{ecrc}

\volume{00}

\firstpage{1}

\journalname{Nuclear Physics B Proceedings Supplement}

\runauth{}

\jid{nuphbp}

\jnltitlelogo{Nuclear Physics B Proceedings Supplement}

\usepackage{amssymb}
\usepackage[figuresright]{rotating}

\newcommand{\bea}{\begin{eqnarray}}
\newcommand{\eea}{\end{eqnarray}}

\begin{document}

\begin{frontmatter}



\dochead{}

%\title{Newtonian versus Relativistic Cosmological Perturbation Theory}
\title{Newtonian, post-Newtonian and Relativistic Cosmological Perturbation Theory}


\author{Jai-chan Hwang${}^{1}$ and Hyerim Noh${}^{2}$}

\address{${}^{1}$Department of Astronomy and Atmospheric Sciences,
                 Kyungpook National University, Daegu, Korea \\
               ${}^{2}$Korea Astronomy and Space Science Institute,
                 Daejon, Korea}

\begin{abstract}
Newtonian cosmological perturbation equations valid to full
nonlinear order are well known in the literature. Assuming the
absence of the transverse-tracefree part of the metric, we present
the general relativistic counterpart valid to full nonlinear order.
The relativistic equations are presented without taking the slicing
(temporal gauge) condition. The equations do have the proper
Newtonian and first post-Newtonian limits. We also present the
relativistic pressure correction terms in the Newtonian hydrodynamic
equations.
\end{abstract}

%\begin{keyword}


%\end{keyword}

\end{frontmatter}
%%%%%%%%%%%%%%%%%%%%%%%%%%%%%%%%%%%%%%%%%%%%%%%%%%%%%%%%%%%%%%%

%%%%%%%%%%%%%%%%%%%%%%%%%%%%%%%%%%%%%%%%%%%%%%%%%%%%%%%%%%%%%%%
%
%  Introduction
%
%%%%%%%%%%%%%%%%%%%%%%%%%%%%%%%%%%%%%%%%%%%%%%%%%%%%%%%%%%%%%%%
\section{Introduction}

Cosmological perturbation theory is an important theoretical tool in
interpreting cosmological observations like the two-dimensional
temperature and polarization anisotropies of the cosmic microwave
background radiation, the three-dimensional distribution and motions
of galaxies, distorted images of galaxies due to gravitational
lensing, etc. The cosmological perturbation equations are well known in the
Newtonian context to fully nonlinear order \cite{Peebles-1980},
whereas the counterparts in Einstein's gravity are known in
linear \cite{Lifshitz-1946,Bardeen-1988} and low-order perturbation
approximation \cite{Bruni-etal-1997}.
Here, we present a self-contained summary of the basic equations of recently formulated fully nonlinear and exact cosmological perturbation theory in Einstein's gravity (Section \ref{sec:NL-eqs}). Comparisons are made with the Newtonian (Sections \ref{sec:Newtonian} and \ref{sec:limit}) and the post-Newtonian equations (Section \ref{sec:PN}). We also present the Newtonian equations in the presence of relativistic pressure (Section \ref{sec:pressure}).

%%%%%%%%%%%%%%%%%%%%%%%%%%%%%%%%%%%%%%%%%%%%%%%%%%%%%%%%%%%%%%%
%
%  Newtonian theory
%
%%%%%%%%%%%%%%%%%%%%%%%%%%%%%%%%%%%%%%%%%%%%%%%%%%%%%%%%%%%%%%%
\section{Newtonian cosmological perturbation theory}
                                      \label{sec:Newtonian}

Newtonian cosmological perturbation equations in the spatially homogeneous and isotropic background world model are \cite{Peebles-1980} \bea
   & & \dot {\widetilde \varrho}
       + 3 {\dot a \over a} \widetilde \varrho
       = - {1 \over a} \nabla \cdot
       \left( \widetilde \varrho {\bf v} \right),
   \label{mass-conservation} \\
   & & \dot {\bf v} + {\dot a \over a} {\bf v}
       + {1 \over a} {\bf v} \cdot \nabla {\bf v}
       = {1 \over a} \nabla U
       - {1 \over a \widetilde \varrho} \nabla \widetilde p,
   \label{momentum-conservation} \\
   & & {\Delta \over a^2} U
       = - 4 \pi G \left( \widetilde \varrho - \varrho \right).
   \label{Poisson-eq}
\eea These are the mass conservation, the momentum conservation, and the
Poisson's equations, respectively; $\widetilde \varrho$, $\widetilde
p$, ${\bf v}$, and $U$ are the mass density, the pressure, the
peculiar velocity, and the perturbed gravitational potential,
respectively; $a(t)$ is the cosmic scale factor. We decompose
the mass density and pressure to the background and perturbed parts
as \bea
   \widetilde \varrho = \varrho + \delta \varrho, \quad
       \widetilde p = p + \delta p.
\eea
Evolution of the background world model is described by equation (\ref{BG-equations}) properly derived in Einstein's gravity.

%%%%%%%%%%%%%%%%%%%%%%%%%%%%%%%%%%%%%%%%%%%%%%%%%%%%%%%%%%%%%%%
%
%  Relativistic theory
%
%%%%%%%%%%%%%%%%%%%%%%%%%%%%%%%%%%%%%%%%%%%%%%%%%%%%%%%%%%%%%%%
\section{General relativistic cosmological perturbation theory}
                                             \label{sec:NL-eqs}

We consider the scalar- and vector-type perturbations in a {\it
flat} background with the metric convention
\cite{Bardeen-1988,Hwang-Noh-2012} \bea
   ds^2 = - \left( 1 + 2 \alpha \right) c^2 d t^2
       - 2 \chi_i c d t d x^i
       + a^2 \left( 1 + 2 \varphi \right) \delta_{ij} d x^i d x^j,
   \label{metric-PT}
\eea where $\alpha$, $\varphi$ and $\chi_i$ are functions of spacetime with {\it arbitrary} amplitudes; index of $\chi_i$ is raised and lowered by $\delta_{ij}$ as the metric.
We {\it ignored} the transverse-tracefree (TT) part of the metric
which is interpreted as the gravitational waves to the linear
perturbation order. The spatial part of the metric is simple
because, in addition to ignoring the TT part, we already have taken
the spatial gauge condition without losing any generality to the
fully nonlinear order \cite{Bardeen-1988,Hwang-Noh-2012}.

We consider a fluid {\it without} anisotropic stress. The energy momentum
tensor is given as \bea
   \widetilde T_{ab}
       = \widetilde \varrho c^2 \widetilde u_a \widetilde u_b
       + \widetilde p \left( \widetilde g_{ab} + \widetilde u_a \widetilde u_b \right),
   \label{Tab}
\eea where tildes indicate the covariant quantities; $\widetilde
u_a$ is the normalized fluid four-vector; $\widetilde \varrho$
includes the internal energy; in explicit presence of the internal
energy we should replace \bea
   \widetilde \varrho
       \rightarrow \widetilde \varrho
       \left( 1 + {1 \over c^2} \widetilde \Pi \right),
   \label{internal-energy}
\eea where $\widetilde \varrho$ in the right-hand-side is the
rest-mass density \cite{Chandrasekhar-1965}. We introduce the following definitions of fluid three-velocities \bea
   \widetilde u_i \equiv a {v_i \over c}, \quad
   {1 \over \widehat \gamma} v_i
       = \widehat v_i
       = {1 \over {\cal N}} \left[ \left( 1 + 2 \varphi \right) \overline v_i
       - {c \over a} \chi_i \right],
   \label{v-relation}
\eea where $\widehat \gamma$ is the Lorentz factor \bea
   \widehat \gamma
       \equiv \sqrt{ 1 + {v^k v_k \over c^2 (1 + 2 \varphi)} }
       = {1 \over \sqrt{ 1
       - {\widehat v^k \widehat v_k \over c^2 (1 + 2 \varphi)}}}
       = {1 \over \sqrt{
       1 - {1 + 2 \varphi \over {\cal N}^2}
       \left( {\overline v^k \over c} - {\chi^k \over a (1 + 2 \varphi)} \right)
       \left( {\overline v_k \over c} - {\chi_k \over a (1 + 2 \varphi)} \right)}},
   \label{Lorentz-factor}
\eea and ${\cal N}$ is related to the lapse function in (\ref{K-bar-eq}). The velocities $\widehat v_i$ and $\overline v_i$ are more physically motivated ones \cite{Hwang-Noh-2012}: $\widehat v_i$ is the fluid three-velocity measured by the Eulerian observer, and $\overline v_i$ is the coordinate three-velocity of fluid; the indices of $v_i$, $\widehat v_i$ and $\overline v_i$ are raised and lowered by $\delta_{ij}$.

We can decompose $\chi_i$ and $\widehat v_i$ into the scalar- and
vector-type perturbations to the nonlinear order as
\cite{Hwang-Noh-2012} \bea
   \chi_i = c \chi_{,i} + \chi_i^{(v)}, \quad
       \widehat v_i \equiv - \widehat v_{,i} + \widehat v_i^{(v)},
\eea with $\chi^{(v)i}_{\;\;\;\;\;,i} \equiv 0 \equiv \widehat
v^{(v)i}_{\;\;\;\;\;,i}$. We assign dimensions as \bea
   & & [a] = [\widetilde g_{ab}] = [\widetilde u_a]
       = [\alpha] = [\varphi] = [\chi^i] = [\widehat v^i/c] = 1,
   \nonumber \\
   & &
       [x^i] = L, \quad
       [\chi] = T, \quad
       [\widehat v/c] = L, \quad
       [\kappa] = T^{-1}, \quad
       [\widetilde T_{ab}] = [\widetilde \varrho c^2]
       = [\widetilde p], \quad
       [G \widetilde \varrho] = T^{-2},
\eea where $\kappa$, the perturbed part of the trace of extrinsic
curvature, will be introduced below.

Here we present the complete set of fully nonlinear perturbation
equations without taking the temporal gauge \cite{Hwang-Noh-2012}.

\noindent The definition of $\kappa$: \bea
   \kappa
       \equiv
       3 {\dot a \over a} \left( 1 - {1 \over {\cal N}} \right)
       - {1 \over {\cal N} (1 + 2 \varphi)}
       \left[ 3 \dot \varphi
       + {c \over a^2} \left( \chi^k_{\;\;,k}
       + {\chi^{k} \varphi_{,k} \over 1 + 2 \varphi} \right)
       \right].
   \label{eq1}
\eea \noindent The ADM energy constraint: \bea
   - {3 \over 2} \left( {\dot a^2 \over a^2}
       - {8 \pi G \over 3} \widetilde \varrho
       - {\Lambda c^2 \over 3} \right)
       + {\dot a \over a} \kappa
       + {c^2 \Delta \varphi \over a^2 (1 + 2 \varphi)^2}
       = {1 \over 6} \kappa^2
       - 4 \pi G \left( \widetilde \varrho + {\widetilde p \over c^2} \right)
       \left( \widehat \gamma^2 - 1 \right)
       + {3 \over 2} {c^2 \varphi^{,i} \varphi_{,i} \over a^2 (1 + 2 \varphi)^3}
       - {c^2 \over 4} \overline{K}^i_j \overline{K}^j_i.
   \label{eq2}
\eea \noindent The ADM momentum constraint: \bea
   & & {2 \over 3} \kappa_{,i}
       + {c \over 2 a^2 {\cal N} ( 1 + 2 \varphi )}
       \left( \Delta \chi_i
       + {1 \over 3} \chi^k_{\;\;,ik} \right)
       + 8 \pi G \left( \widetilde \varrho + {\widetilde p \over c^2} \right)
       a \widehat \gamma^2 {\widehat v_{i} \over c^2}
   \nonumber \\
   & & \qquad
       =
       {c \over a^2 {\cal N} ( 1 + 2 \varphi)}
       \Bigg\{
       \left( {{\cal N}_{,j} \over {\cal N}}
       - {\varphi_{,j} \over 1 + 2 \varphi} \right)
       \left[ {1 \over 2} \left( \chi^{j}_{\;\;,i} + \chi_i^{\;,j} \right)
       - {1 \over 3} \delta^j_i \chi^k_{\;\;,k} \right]
   \nonumber \\
   & & \qquad
       - {\varphi^{,j} \over (1 + 2 \varphi)^2}
       \left( \chi_{i} \varphi_{,j}
       + {1 \over 3} \chi_{j} \varphi_{,i} \right)
       + {{\cal N} \over 1 + 2 \varphi} \nabla_j
       \left[ {1 \over {\cal N}} \left(
       \chi^{j} \varphi_{,i}
       + \chi_{i} \varphi^{,j}
       - {2 \over 3} \delta^j_i \chi^{k} \varphi_{,k} \right) \right]
       \Bigg\}.
   \label{eq3}
\eea \noindent The trace of ADM propagation: \bea
   & & - 3 {1 \over {\cal N}}
       \left( {\dot a \over a} \right)^{\displaystyle\cdot}
       - 3 {\dot a^2 \over a^2}
       - 4 \pi G \left( \widetilde \varrho + 3 {\widetilde p \over c^2} \right)
       + \Lambda c^2
       + {1 \over {\cal {\cal N}}} \dot \kappa
       + 2 {\dot a \over a} \kappa
       + {c^2 \Delta {\cal N} \over a^2 {\cal N} (1 + 2 \varphi)}
   \nonumber \\
   & & \qquad
       = {1 \over 3} \kappa^2
       + 8 \pi G \left( \widetilde \varrho + {\widetilde p \over c^2} \right)
       \left( \widehat \gamma^2 - 1 \right)
       - {c \over a^2 {\cal N} (1 + 2 \varphi)} \left(
       \chi^{i} \kappa_{,i}
       + c {\varphi^{,i} {\cal N}_{,i} \over 1 + 2 \varphi} \right)
       + c^2 \overline{K}^i_j \overline{K}^j_i.
   \label{eq4}
\eea \noindent The tracefree ADM propagation: \bea
   & & \left( {1 \over {\cal N}} {\partial \over \partial t}
       + 3 {\dot a \over a}
       - \kappa
       + {c \chi^{k} \over a^2 {\cal N} (1 + 2 \varphi)} \nabla_k \right)
       \Bigg\{ {c \over a^2 {\cal N} (1 + 2 \varphi)}
       \Bigg[
       {1 \over 2} \left( \chi^i_{\;\;,j} + \chi_j^{\;,i} \right)
       - {1 \over 3} \delta^i_j \chi^k_{\;\;,k}
       - {1 \over 1 + 2 \varphi}
   \nonumber \\
   & & \qquad
       \times
       \left( \chi^{i} \varphi_{,j}
       + \chi_{j} \varphi^{,i}
       - {2 \over 3} \delta^i_j \chi^{k} \varphi_{,k} \right)
       \Bigg] \Bigg\}
       - {c^2 \over a^2 ( 1 + 2 \varphi)}
       \left[ {1 \over 1 + 2 \varphi}
       \left( \nabla^i \nabla_j - {1 \over 3} \delta^i_j \Delta \right) \varphi
       + {1 \over {\cal N}}
       \left( \nabla^i \nabla_j - {1 \over 3} \delta^i_j \Delta \right) {\cal N} \right]
   \nonumber \\
   & & \qquad
       =
       8 \pi G \left( \widetilde \varrho + {\widetilde p \over c^2} \right)
       \left[ {\widehat \gamma^2 \widehat v^i \widehat v_j \over c^2 (1 + 2 \varphi)}
       - {1 \over 3} \delta^i_j \left( \widehat \gamma^2 - 1 \right)
       \right]
       + {c^2 \over a^4 {\cal N}^2 (1 + 2 \varphi)^2}
       \Bigg[
       {1 \over 2} \left( \chi^{i,k} \chi_{j,k}
       - \chi_{k,j} \chi^{k,i} \right)
   \nonumber \\
   & & \qquad
       + {1 \over 1 + 2 \varphi} \left(
       \chi^{k,i} \chi_k \varphi_{,j}
       - \chi^{i,k} \chi_j \varphi_{,k}
       + \chi_{k,j} \chi^k \varphi^{,i}
       - \chi_{j,k} \chi^i \varphi^{,k} \right)
       + {2 \over (1 + 2 \varphi)^2} \left(
       \chi^{i} \chi_{j} \varphi^{,k} \varphi_{,k}
       - \chi^{k} \chi_{k} \varphi^{,i} \varphi_{,j} \right) \Bigg]
   \nonumber \\
   & & \qquad
       - {c^2 \over a^2 (1 + 2 \varphi)^2}
       \Bigg[ {3 \over 1 + 2 \varphi}
       \left( \varphi^{,i} \varphi_{,j}
       - {1 \over 3} \delta^i_j \varphi^{,k} \varphi_{,k} \right)
       + {1 \over {\cal N}} \left(
       \varphi^{,i} {\cal N}_{,j}
       + \varphi_{,j} {\cal N}^{,i}
       - {2 \over 3} \delta^i_j \varphi^{,k} {\cal N}_{,k} \right) \Bigg].
   \label{eq5}
\eea \noindent The covariant energy conservation: \bea
   & &
       \left[ {\partial \over \partial t}
       + {1 \over a ( 1 + 2 \varphi )} \left( {\cal N} \widehat v^k
       + {c \over a} \chi^k \right) \nabla_k \right] \widetilde \varrho
       + \left( \widetilde \varrho + {\widetilde p \over c^2} \right)
       \Bigg\{
       {\cal N} \left( 3 {\dot a \over a} - \kappa \right)
   \nonumber \\
   & & \qquad
       +
       {({\cal N} \widehat v^k)_{,k} \over a (1 + 2 \varphi)}
       + {{\cal N} \widehat v^k \varphi_{,k} \over a (1 + 2 \varphi)^2}
       + {1 \over \widehat \gamma}
       \left[ {\partial \over \partial t}
       + {1 \over a ( 1 + 2 \varphi )} \left( {\cal N} \widehat v^k
       + {c \over a} \chi^k \right) \nabla_k \right] \widehat \gamma \Bigg\}
       = 0.
   \label{eq6}
\eea \noindent The covariant momentum conservation: \bea
   & & {1 \over a \widehat \gamma}
       \left[ {\partial \over \partial t}
       + {1 \over a ( 1 + 2 \varphi )} \left( {\cal N} \widehat v^k
       + {c \over a} \chi^k \right) \nabla_k \right]
       \left( a \widehat \gamma \widehat v_i \right)
       + \widehat v^k \nabla_i \left( {c \chi_k \over a^2 ( 1 + 2
       \varphi)} \right)
       + {c^2 \over a} {\cal N}_{,i}
       - \left( 1 - {1 \over \widehat \gamma^2} \right) {c^2 {\cal N}
       \varphi_{,i} \over a (1 + 2 \varphi)}
   \nonumber \\
   & & \qquad
       + {1 \over \widetilde \varrho + {\widetilde p \over c^2}}
       \left\{
       {{\cal N} \over a \widehat \gamma^2} \widetilde p_{,i}
       + {\widehat v_i \over c^2}
       \left[ {\partial \over \partial t}
       + {1 \over a ( 1 + 2 \varphi )} \left( {\cal N} \widehat v^k
       + {c \over a} \chi^k \right) \nabla_k \right] \widetilde p \right\}
       = 0,
   \label{eq7}
\eea where $\overline{K}^i_j$ and $N$ are the tracefree part of the extrinsic
curvature and the lapse function, respectively, with \bea
   & & \overline{K}^i_j \overline{K}^j_i
       = {1 \over a^4 {\cal N}^2 (1 + 2 \varphi)^2}
       \Bigg\{
       {1 \over 2} \chi^{i,j} \left( \chi_{i,j} + \chi_{j,i} \right)
       - {1 \over 3} \chi^i_{\;\;,i} \chi^j_{\;\;,j}
       - {4 \over 1 + 2 \varphi} \left[
       {1 \over 2} \chi^i \varphi^{,j} \left(
       \chi_{i,j} + \chi_{j,i} \right)
       - {1 \over 3} \chi^i_{\;\;,i} \chi^j \varphi_{,j} \right]
   \nonumber \\
   & & \qquad
       + {2 \over (1 + 2 \varphi)^2} \left(
       \chi^{i} \chi_{i} \varphi^{,j} \varphi_{,j}
       + {1 \over 3} \chi^i \chi^j \varphi_{,i} \varphi_{,j} \right) \Bigg\}, \quad
       {\cal N} \equiv \sqrt{ 1 + 2 \alpha
       + {\chi^k \chi_k \over a^2 ( 1 + 2 \varphi )}}.
   \label{K-bar-eq}
\eea

Apparently the basic set of perturbation equations in Einstein's
gravity in (\ref{eq1})-(\ref{eq7}) looks quite complicated
compared with the Newtonian ones in (\ref{mass-conservation})-(\ref{Poisson-eq}). In fact equations
(\ref{eq1})-(\ref{eq7}) are redundant; for example, the Einstein's
equations imply the conservation equations. More importantly, in
the above equations, we have not taken the temporal gauge
condition yet. Equations (\ref{eq1})-(\ref{eq7}) are presented without taking the temporal gauge (hypersurface or slicing) condition. As the temporal
gauge condition we can impose any one of the following conditions
\bea
   & & {\rm comoving \; gauge:}              \hskip 1.89cm     \widehat v \equiv 0,
   \nonumber \\
   & & {\rm zero\!-\!shear \; gauge:}        \hskip 1.67cm     \chi \equiv 0,
   \nonumber \\
   & & {\rm uniform\!-\!curvature \; gauge:} \hskip .55cm      \varphi \equiv 0,
   \nonumber \\
   & & {\rm uniform\!-\!expansion \; gauge:} \hskip .47cm      \kappa \equiv 0,
   \nonumber \\
   & & {\rm uniform\!-\!density \; gauge:}   \hskip .88cm      \delta \equiv 0,
   \label{temporal-gauges-NL}
\eea or combinations of these to each perturbation order; we may
call these the fundamental gauge conditions. With the imposition of
any of these temporal gauge conditions the remaining perturbation
variables are free from the remnant (spatial and temporal) gauge
mode, and have unique gauge-invariant combinations. Thus, we can
regard each perturbation variable in those gauges as the
gauge-invariant one to nonlinear order
\cite{Bardeen-1988,Hwang-Noh-2012}.

To the background order, (\ref{eq2}), (\ref{eq4}) and (\ref{eq6}), respectively, give \bea
   {\dot a^2 \over a^2} = {8 \pi G \over 3} \varrho
       + {\Lambda c^2 \over 3}, \quad
       {\ddot a \over a} = - {4 \pi G \over 3}
       \left( \varrho + 3 {p \over c^2} \right) + {\Lambda c^2 \over 3}, \quad
       \dot \varrho + 3 {\dot a \over a}
       \left( \varrho + {p \over c^2} \right) = 0,
   \label{BG-equations}
\eea where $\Lambda$ is the cosmological constant. In the Newtonian
limit we ignore $p$ compared with $\varrho c^2$.

%%%%%%%%%%%%%%%%%%%%%%%%%%%%%%%%%%%%%%%%%%%%%%%%%%%%%%%%%%%%%%%
%
%  Newtonian limit
%
%%%%%%%%%%%%%%%%%%%%%%%%%%%%%%%%%%%%%%%%%%%%%%%%%%%%%%%%%%%%%%%
\section{Newtonian limit}
                                          \label{sec:limit}

The infinite speed-of-light limit leads to the Newtonian equations
\cite{Chandrasekhar-1965}. That is, as the
Newtonian limit we consider the weak-gravity, the slow-motion,
negligible pressure and internal energy compared with the energy density, and the small-scale
(subhorizon) limits \bea
   \alpha \ll 1, \quad
       \varphi \ll 1, \quad
       {\widehat v^k \widehat v_k \over c^2} \ll 1, \quad
       \widetilde p \ll \widetilde \varrho c^2, \quad
       {1 \over c^2} \widetilde \Pi \ll 1, \quad
       {c^2 k^2 \over a^2 H^2} \gg 1,
   \label{NL-limit}
\eea where $k$ the comoving wave-number with $\Delta = - k^2$; $H
\equiv \dot a/a$; in the presence of the cosmological constant
$\Lambda$, we consider $H^2 \sim 8 \pi G \varrho$. We identify \bea
   \alpha = - {1 \over c^2} U, \quad
       \varphi = {1 \over c^2} V, \quad
       \widehat v^k = {\bf v},
   \label{identification}
\eea where $U$ and $V$ correspond to the Newtonian and the post-Newtonian perturbed gravitational potentials, respectively
\cite{Chandrasekhar-1965,Hwang-etal-2008}; equation (\ref{eq5})
gives $\varphi = - \alpha$, thus $V = U$. We can show that equations
(\ref{mass-conservation})-(\ref{Poisson-eq}) follow from equations
(\ref{eq6}), (\ref{eq7}), and (\ref{eq4}), respectively, in both the
zero-shear gauge ($\chi \equiv 0$) and the uniform-expansion gauge
($\kappa \equiv 0$).

%%%%%%%%%%%%%%%%%%%%%%%%%%%%%%%%%%%%%%%%%%%%%%%%%%%%%%%%%%%%%%%
%
%  1PN
%
%%%%%%%%%%%%%%%%%%%%%%%%%%%%%%%%%%%%%%%%%%%%%%%%%%%%%%%%%%%%%%%
\section{First post-Newtonian approximation}
                                          \label{sec:PN}

To the first post-Newtonian (1PN) order our metric convention is
\cite{Chandrasekhar-1965,Hwang-etal-2008} \bea
   ds^2 = - \left[ 1 - {1 \over c^2} 2 U
       + {1 \over c^4} \left( 2 U^2 - 4 \Phi \right) \right] c^2 d t^2
       - {1 \over c^3} 2 a P_i c dt d x^i
       + a^2 \left( 1 + {1 \over c^2} 2 V \right) \delta_{ij} d x^i d x^j,
   \label{metric-PN}
\eea where the index of $P_i$ is raised and lowered by
$\delta_{ij}$. Similarly as in the metric of perturbation theory in equation (\ref{metric-PT}), in the spatial part of the metric we
already have taken spatial gauge conditions without losing any
generality, and have ignored the TT-type perturbation
\cite{Hwang-etal-2008}. The dimensions are the following \bea
   [U] = [V] = c^2, \quad
       [P_i] = c^3, \quad
       [\Phi] = c^4.
\eea Comparing equation (\ref{metric-PT}) in the perturbation
theory with equation (\ref{metric-PN}) in the PN approach, to the
1PN order we have  \bea
   \alpha = - {1 \over c^2} \left[ U
       - {1 \over c^2} \left( U^2 - 2 \Phi \right) \right], \quad
       \varphi = {1 \over c^2} V, \quad
       \kappa = - {1 \over c^2} \left( 3 {\dot a \over a} U
       + 3 \dot V
       + {1 \over a} P^k_{\;\;,k} \right), \quad
       \chi_i = {1 \over c^3} a P_i.
   \label{PT-PN}
\eea

Using the identifications between the two approaches made in equation (\ref{PT-PN}), we can derive the 1PN equations from the perturbation equations in (\ref{eq1})-(\ref{eq7}); $\overline v_i$ is used as the fluid three-velocity in \cite{Hwang-etal-2008}.
Equation (\ref{eq5}) to 1PN order gives $V = U$. From equations (\ref{eq6}), (\ref{eq7}), (\ref{eq4}) and (\ref{eq3}), respectively,
we can show \bea
   & &
       {1 \over a^3} \left( a^3 \widetilde \varrho \right)^{\displaystyle\cdot}
       + {1 \over a} \left( \widetilde \varrho \overline{v}^i \right)_{,i}
       = - {1 \over c^2} \left[ \widetilde \varrho \left( {\partial \over \partial t}
       + {1 \over a} \overline{\bf v} \cdot \nabla \right)
       \left( {1 \over 2} \overline{v}^2 + 3 U + \widetilde \Pi \right)
       + \left( 3 {\dot a \over a}
       + {1 \over a} \nabla \cdot \overline{\bf v} \right) \widetilde p \right],
   \label{E-conserv-PN} \\
   & &
       {1 \over a} \left( a \overline{v}_i \right)^{\displaystyle\cdot}
       + {1 \over a} \overline{v}_{i,k} \overline{v}^k
       - {1 \over a} U_{,i}
       + {1 \over a} {\widetilde p_{,i} \over \widetilde \varrho}
       = {1 \over c^2} \Bigg[
       {1 \over a} \overline{v}^2 U_{,i}
       + {2 \over a} \left( \Phi - U^2 \right)_{,i}
       + {1 \over a} \left( a P_i \right)^{\displaystyle\cdot}
       + {1 \over a} \overline{v}^k \left( P_{i,k} - P_{k,i} \right)
   \nonumber \\
   & & \qquad
       + {1 \over a} \left( \overline{v}^2 + 4 U
       + \widetilde \Pi + {\widetilde p \over \widetilde \varrho}
       \right) {\widetilde p_{,i} \over \widetilde \varrho}
       - \overline{v}_i \left( {\partial \over \partial t}
       + {1 \over a} \overline{\bf v} \cdot \nabla \right)
       \left( {1 \over 2} \overline{v}^2 + 3 U \right)
       - \overline{v}_i {1 \over \widetilde \varrho} \left( {\partial \over \partial t}
       + {1 \over a} \overline{\bf v} \cdot \nabla \right) \widetilde p
       \Bigg],
   \label{Mom-conserv-PN} \\
   & &
       {\Delta \over a^2} U
       + 4 \pi G \left( \widetilde \varrho - \varrho \right)
       = - {1 \over c^2} \Bigg\{
       {1 \over a^2} \left[
       2 \Delta \Phi
       - 2 U \Delta U
       + \left( a P^i_{\;\; ,i} \right)^{\displaystyle\cdot}
       \right]
       + 3 \ddot U
       + 9 {\dot a \over a} \dot U
       + 6 {\ddot a \over a} U
   \nonumber \\
   & & \qquad
       + 8 \pi G \left[ \widetilde \varrho \overline{v}^2
       + {1 \over 2} \left( \widetilde \varrho \widetilde \Pi
       - \varrho \Pi \right)
       + {3 \over 2} \left( \widetilde p - p \right) \right]
       \Bigg\},
   \label{Raychaudhury-eq} \\
   & &
       0 = {1 \over a^2} \left( P^k_{\;\; ,ki}
       - \Delta P_i \right)
       - 16 \pi G \widetilde \varrho \overline{v}_i
       + {4 \over a} \left( \dot U + {\dot a \over a} U
       \right)_{,i},
   \label{Mom-constr-PN}
\eea where the right-hand-sides are 1PN order, and we have recovered
$\widetilde \Pi$ explicitly; $\overline v^2 \equiv \overline v^k \overline v_k$. These are the same as cosmological 1PN
equations in \cite{Chandrasekhar-1965,Hwang-etal-2008}. To 1PN order, from equation (\ref{v-relation}) we have \bea
   \overline v_i
       = \left( 1 - {3 \over c^2} U \right) \widehat v_i
       + {1 \over c^2} P_i, \quad
       \widehat v_i = \left( 1 - {v^2 \over 2 c^2} \right) v_i,
\eea where $v^2 \equiv v^k v_k$. Notice that in order to derive the 1PN equations we have not imposed the temporal gauge condition; in the PN expansion $\chi_i$ in equation (\ref{PT-PN}) is already higher order.
In the above equations we can still impose a gauge condition  \cite{Hwang-etal-2008} \bea
   {1 \over a} P^i_{\;\; ,i} + n \dot U + m {\dot a \over a} U = 0,
   \label{Gauge-PN}
\eea where $n$ and $m$ can be arbitrary real numbers. The
uniform-expansion gauge takes $n = 3 = m$, and the transverse-shear
gauge takes $n = 0 = m$. We note that the comoving gauge, the uniform-curvature gauge and the uniform-density gauge are not available in the PN approximation.

%%%%%%%%%%%%%%%%%%%%%%%%%%%%%%%%%%%%%%%%%%%%%%%%%%%%%%%%%%%%%%%
%
%  Pressure
%
%%%%%%%%%%%%%%%%%%%%%%%%%%%%%%%%%%%%%%%%%%%%%%%%%%%%%%%%%%%%%%%
\section{Relativistic pressure corrections}
                                      \label{sec:pressure}

We can derive relativistic pressure correction terms in equations
(\ref{mass-conservation})-(\ref{Poisson-eq}) by relaxing the
conditions $\widetilde p \ll \widetilde \varrho c^2$ and $\widetilde
\Pi/c^2 \ll 1$ in equation (\ref{NL-limit}). We take the zero-shear
gauge by setting $\chi \equiv 0$. From equations (\ref{eq6}),
(\ref{eq7}), and (\ref{eq4}), respectively, we have \bea
   & & \dot {\widetilde \varrho}
         + 3 {\dot a \over a} \left( \widetilde \varrho
         + {\widetilde p \over c^2} \right)
         + {1 \over a} \nabla \cdot
       \left[ \left( \widetilde \varrho
         + {\widetilde p \over c^2} \right) {\bf v} \right]
       = {1 \over c^2} {2 \over a}
       {\bf v} \cdot \nabla \widetilde p,
   \label{mass-conservation-p} \\
   & & \dot {\bf v} + {\dot a \over a} {\bf v}
       + {1 \over a} {\bf v} \cdot \nabla {\bf v}
       - {1 \over a} \nabla U
       = - {1 \over \widetilde \varrho + \widetilde p/c^2}
       \left( {1 \over a} \nabla \widetilde p
       + {\dot {\widetilde p} \over c^2} {\bf v} \right),
   \label{momentum-conservation-p} \\
   & & {\Delta \over a^2} U
       = - 4 \pi G \left( \widetilde \varrho - \varrho \right),
   \label{Poisson-eq-p}
\eea where $\widetilde \varrho$ now includes the internal energy. We
note that the pressure corrections in the above equations differ
from the known ones in the literature \cite{Harrison-1965}. The term
in the right-hand-side of equation (\ref{mass-conservation-p}) was
not known in the literature. Notice, in particular, the absence of
pressure correction term in equation (\ref{Poisson-eq-p}). The
presence of a pressure correction term, $- 12 \pi G (\widetilde p -
p)$, in the right-hand-side of equation (\ref{Poisson-eq-p}) was
often suggested in the literature
\cite{Whittaker-1935,Harrison-1965}.

%%%%%%%%%%%%%%%%%%%%%%%%%%%%%%%%%%%%%%%%%%%%%%%%%%%%%%%%%%%%%%%
%
%  Discussion
%
%%%%%%%%%%%%%%%%%%%%%%%%%%%%%%%%%%%%%%%%%%%%%%%%%%%%%%%%%%%%%%%
\section{Discussion}
                                       \label{sec:Discussion}

The Newtonian equations in (\ref{mass-conservation})-(\ref{Poisson-eq}) are derived as the $1/c \rightarrow 0$ limit of Einstein's theory in the zero-shear gauge and the uniform-expansion gauge.
The 1PN equations in (\ref{E-conserv-PN})-(\ref{Mom-constr-PN}) are derived as a weak gravity limit keeping $c^{-2}$-order terms compared with the Newtonian limit.
Both the Newtonian and 1PN equations are correctly recovered from our fully nonlinear and exact formulation of cosmological perturbation in Einstein's gravity in equations (\ref{eq1})-(\ref{eq7}).
In all (Newtonian, 1PN, and Einstein's) cases the equations are fully nonlinear and exact.
In the nonlinear cosmological perturbation equations in the Einstein's gravity and in the 1PN approximation we can still impose one temporal gauge condition suggested in equations (\ref{temporal-gauges-NL}) and (\ref{Gauge-PN}), respectively.

As we consider nonlinear perturbations, ignoring the TT part can be
regarded as a serious assumption restricting the range of
applications of our nonlinear perturbation formulation in Einstein's
gravity. Except for this shortcoming, as we have not decomposed the
background parts, the basic set in equations (\ref{eq1})-(\ref{eq7})
can also be regarded as an exact one; by contrast, except for the
mass conservation equation, the equations in the Newtonian and 1PN
approximation are valid only to the perturbed variables in
cosmology; the cosmological background equations are subtracted by
using the equation derived in Einstein's gravity
\cite{Hwang-etal-2008}. Meanwhile, by setting the background order
quantities as $a \equiv 1$ and $\varrho \equiv 0 \equiv p$,
equations (\ref{mass-conservation})-(\ref{Poisson-eq}) and equations
(\ref{E-conserv-PN})-(\ref{Mom-constr-PN}) are valid for the
Newtonian and 1PN hydrodynamic equations, respectively, in the
Minkowski background \cite{Peebles-1980,Chandrasekhar-1965}.

Derivation of the equations in these proceedings with some applications will be presented in \cite{Hwang-Noh-2012,Hwang-Noh-2012-Newtonian}.

%%%%%%%%%%%%%%%%%%%%%%%%%%%%%%%%%%%%%%%%%%%%%%%%%%%%%%%%%%%%%%%
%
% Acknowledgments
%
%%%%%%%%%%%%%%%%%%%%%%%%%%%%%%%%%%%%%%%%%%%%%%%%%%%%%%%%%%%%%%%
\section*{Acknowledgments}

H.N.\ was supported by grant No.\ 2012 R1A1A2038497 from NRF. J.H.\
was supported by KRF Grant funded by the Korean Government
(KRF-2008-341-C00022).

%%%%%%%%%%%%%%%%%%%%%%%%%%%%%%%%%%%%%%%%%%%%%%%%%%%%%%%%%%%%%%%%%
%
%   References
%
%%%%%%%%%%%%%%%%%%%%%%%%%%%%%%%%%%%%%%%%%%%%%%%%%%%%%%%%%%%%%%%%%

%\nocite{*}
\bibliographystyle{elsarticle-num}
%\bibliography{martin}

%   References
%
%%%%%%%%%%%%%%%%%%%%%%%%%%%%%%%%%%%%%%%%%%%%%%%%%%%%%%%%%%%%%%%%%

%%%%%%%%%%%%%%%%%%%%%%%%%%%%%%%%%%%%%%%%%%%%%%%%%%%%%%%%%%%%%%%

%%%%%%%%%%%%%%%%%%%%%%%%%%%%%%%%%%%%%%%%%%%%%%%%%%%%%%%%%%%%%%%
\end{document}